\documentclass[conference]{IEEEtran}
\IEEEoverridecommandlockouts
\usepackage{cite}
\usepackage{amsmath,amssymb,amsfonts}
\usepackage{algorithmic}
\usepackage{graphicx}
\usepackage{placeins}

\usepackage{textcomp}
\usepackage{xcolor}
\usepackage{subcaption}

\def\BibTeX{{\rm B\kern-.05em{\sc i\kern-.025em b}\kern-.08em
    T\kern-.1667em\lower.7ex\hbox{E}\kern-.125emX}}
\begin{document}

\title{Addressing deepfake issue in selfie-banking through camera-based authentication\\
}

\author{\IEEEauthorblockN{Subhrojyoti Mukherjee}
\IEEEauthorblockA{\textit{Metallurgical And Materials Engineering} \\
\textit{Indian Institute Of Technology}\\
Kharagpur, India \\
subhrojm@kgpian.iitkgp.ac.in}
\and
\IEEEauthorblockN{Manoranjan Mohanty}
\IEEEauthorblockA{\textit{Information Systems} \\
\textit{Carnegie Mellon University}\\
Doha, Qatar \\
mmohanty@cmu.edu}

}

\maketitle

\begin{abstract}
  
Fake images in selfie banking are increasingly becoming a threat. Previously, it was just Photoshop, but now deep learning technologies enable us to create highly realistic fake identities, which fraudsters exploit to bypass biometric systems such as facial recognition in online banking. This paper explores the use of an already established forensic recognition system, previously used for picture camera localization, in deepfake detection.
\end{abstract}


\section{Introduction}
Facial recognition is a popular form of authentication in the banking and financial industry due to its user friendliness \cite{Vamshi_Chandana_Priya_Ramya_Rao_2024}. The banking using this form of authentication is also called \textit{ selfi-banking}. In face-based authentication, a customer is authenticated using her live face, e.g., taken as a live photo or video. Selfie banking has become popular because it does not require one to remember a password, and hence makes the login and logout of customers simple. 

Liveness detection has been used to deal with a possible spoofing attack in facial recognition \cite {articlea}. The popular passive form of liveness detection analyses the face in real-time to find features, such as texture, depth cues, etc.,  that are only possible in a live real face but not in a spoof\cite{articlea}. Liveness detection is the reason why our photos or videos cannot replace us when authentication is needed through our face.  

\begin{figure}[htbp]
 \centering
\includegraphics[width=0.5\textwidth]{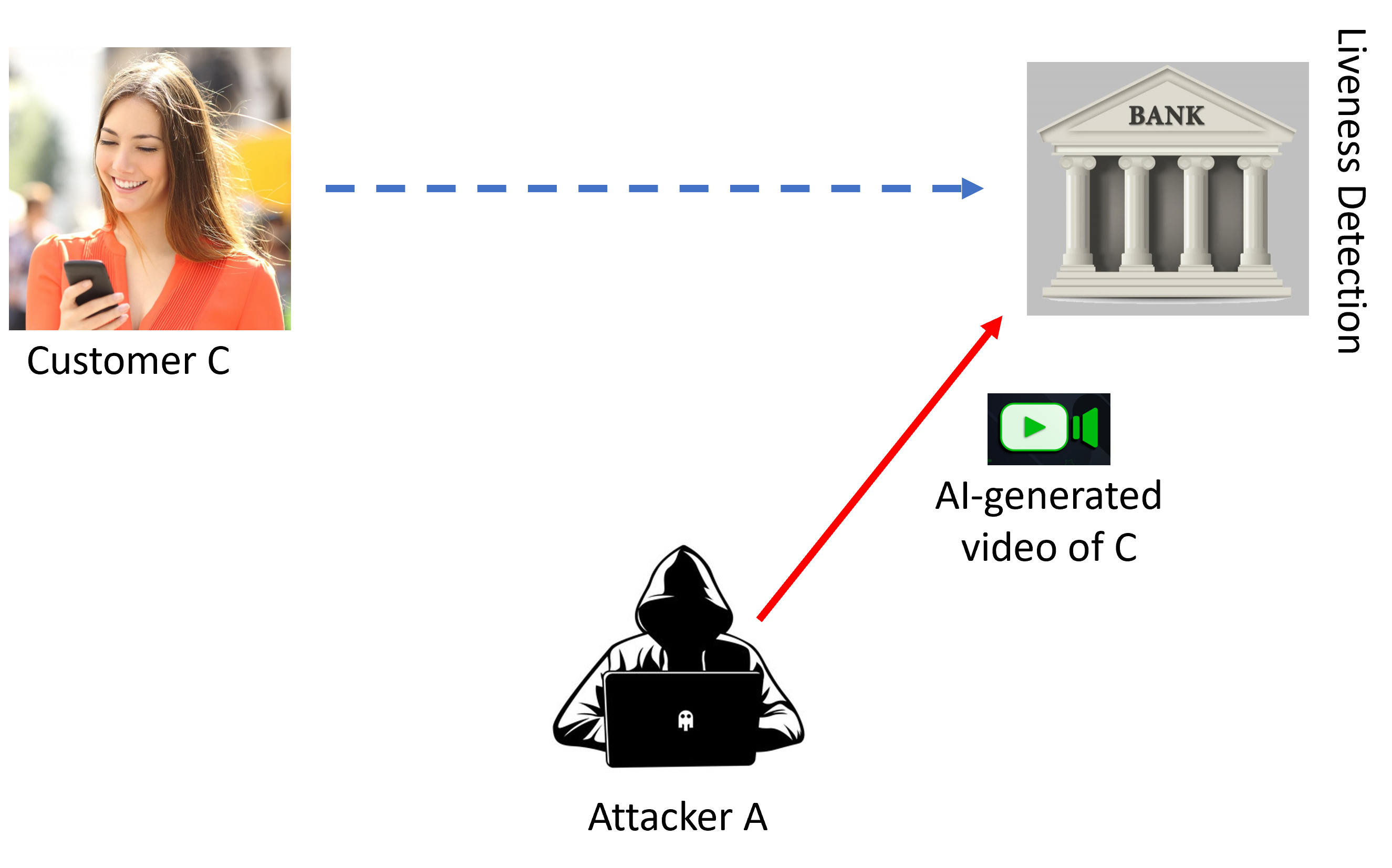}
\caption{Deepfake a problem for selfie-banking}
\label{fig1}
\end{figure}

However, the widely available deepfake technology creates a challenge for liveness detection, as features that were only possible in live faces can now be part of a deepfake. This implies that the old single factor authentication scheme is now in danger. Creating a deepfake is becoming very easy. Deepfake can now be created using a single photo of a victim (e.g., a profile photo). Face swap tools like \textit{Akool, Face Swapper, and pc ai}, are readily available for anyone to use. As a result, the single-factor authentication-based selfie banking system can be spoofed. Figure~\ref{fig1} shows a scenario in which an attacker can log into a customer's bank account using a deep fake of the customer. 

\begin{figure}[htbp]
 \centering
\includegraphics[width=0.5\textwidth]{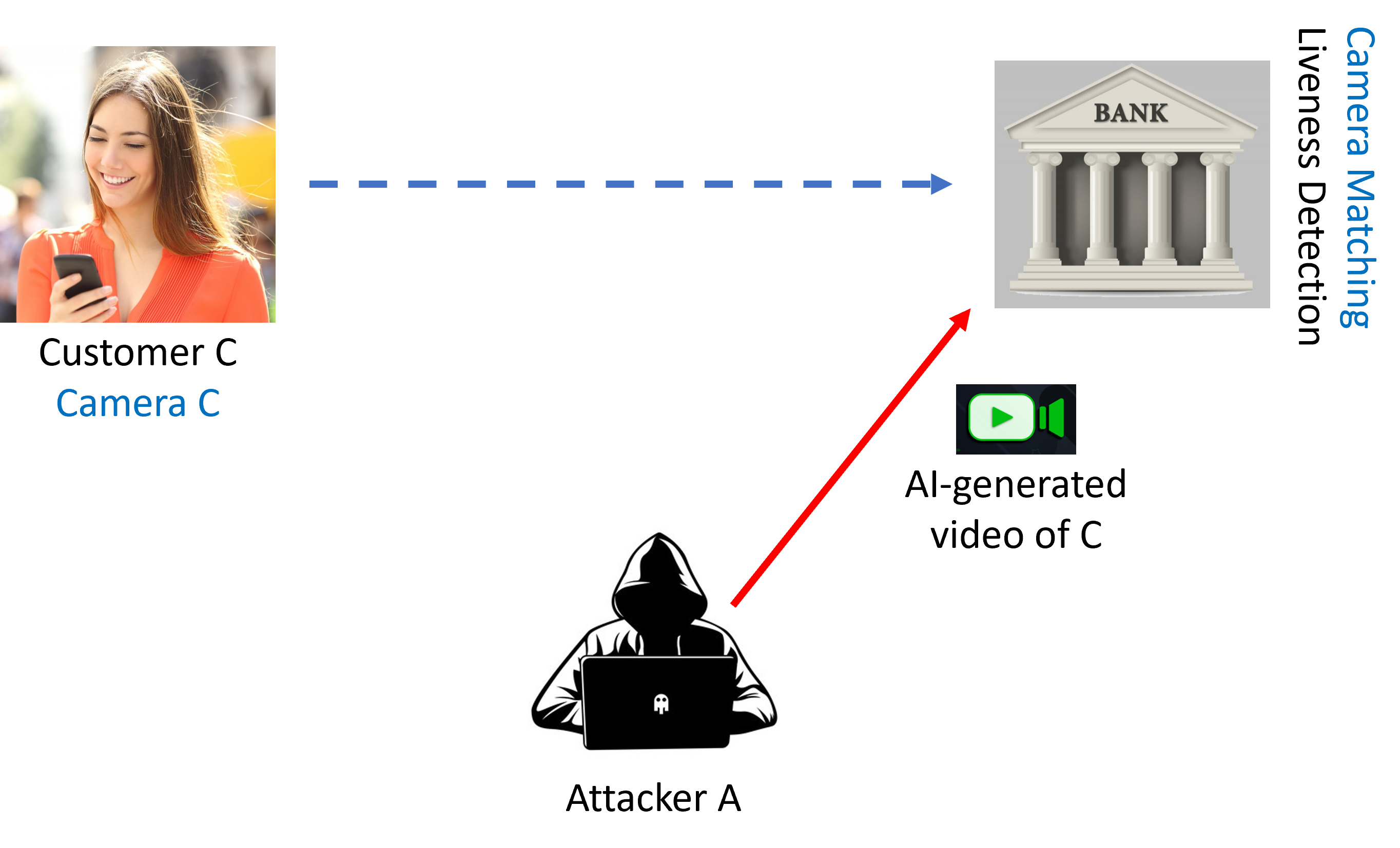}
\caption{Our Appraoch}
\label{fig2}
\end{figure}

In this paper, we address this vulnerability in facial recognition by proposing source camera authentication as a second-factor authentication (Figure~\ref{fig2}). Camera authentication is done using the time-tested PRNU (Photo Response Non-Uniformity) \cite{10.1117/12.805701} method, which can perform automatic authentication without requiring any intervention from a user. First, a camera fingerprint is computed from a customer's camera and stored on bank's server when the user first registers herself in the bank. This registration process also need to be repeated when the customer uses a new camera (e.g., uses a new smartphone) for the first time. The fingerprint is computed from around $20–25$ images from the customer's camera; physical access to the camera is not needed. When the user tries to authenticate herself using her face, the facial features is sent to the bank for facial recognition. At the same time, the camera fingerprint computed from the fresh set of images (clicked during the facial recognition) is also sent to the bank. This new fingerprint is then matched to the stored fingerprint to see if both belong to the same camera. If so, the camera is authenticated. if the facial authentication was also a success, the user is authenticated.

The PRNU-based method has been shown to be robust against a number of scenarios and attacks \cite{10.1117/12.805701}. It is difficult to plant PRNU on an image without access to the camera \cite{10.1117/12.805701}. Therefore, the above authentication scheme is well poised to deal with the man-in-the middle attack where the middle man tries to fool the authentication system using a fake AI-generated image or video (i.e., deepfake). The PRNU also withstands compression, image processing, and cannot be removed from an image \cite{10.1117/12.805701}.

The experiment with a laptop camera, a smartphone, and a webcam shows the usefulness of our method. A deepfake video generated using a photo from these cameras defeated a liveness detection scheme. The Akool software was used to generate the deepfake, and an internal code was developed for liveness detection. However, the deepfake failed the PRNU test because the fingerprint obtained from the deepfake did not match the camera fingerprint.

\section{Related Works} \label{sec:releate}

Deepfake detection is an active area of research. Early approaches relied on identifying hand-crafted artifacts and physiological inconsistencies in manipulated videos, such as irregular blinking, abnormal head poses, and inconsistent facial landmarks, as well as traditional image forensics techniques like JPEG compression and frequency domain analysis~\cite{li2018ictu, matern2019exploiting, frank2020leveraging}. While initially effective, these methods struggled against more sophisticated, adversarially-trained deepfakes. As realism increased, the field shifted toward data-driven techniques~\cite{romeo2024fasterliesrealtimedeepfake}, with deep learning models such as Convolutional Neural Networks (CNNs) and Recurrent Neural Networks (RNNs) being used to capture spatial and temporal artifacts~\cite{rossler2019faceforensics++}\cite{10057390}. Multimodal approaches incorporating visual, audio, and physiological cues—like lip-sync accuracy and remote photoplethysmography (rPPG)—have also emerged~\cite{ciftci2020fakecatcher}. Large-scale datasets such as FaceForensics++ and the DeepFake Detection Challenge (DFDC) have further driven progress by providing diverse training material~\cite{dolhansky2020deepfake}. Nonetheless, generalizing to unseen data and ensuring robustness in real-world scenarios remain persistent challenges.


Photo-Response Non-Uniformity (PRNU) analysis has emerged as a promising deepfake detection technique by leveraging the unique sensor noise patterns introduced by camera hardware. Unlike methods that focus on visual artifacts or deep learning features, PRNU detects inconsistencies or absence of this intrinsic noise in synthetic content~\cite{verdoliva2020media}. Cozzolino and Verdoliva~\cite{cozzolino2019noiseprint} proposed Noiseprint, a CNN-based approach for extracting camera fingerprints, which has been adapted to spot GAN-generated images through spatial noise discrepancies. While effective even in high-quality deepfakes, PRNU methods face limitations with low-resolution or compressed media, and evolving generative models may attempt to replicate sensor noise to bypass detection.

PRNU is a unique noise pattern caused by imperfections in a camera’s sensor, consistently embedded in images and videos, enabling device identification~\cite{10.1117/12.805701}. Video source attribution involves computing a camera fingerprint from a known video and comparing it with one from an anonymous video. Fingerprints are extracted by estimating PRNU noise from individual frames using filters like Wavelet, then combining the noise to form a single signature, with matching done via the PCE method—where a score of 50 or above indicates a match. Recent work has enhanced PRNU robustness against compression, resizing, stabilization, and other distortions~\cite{mandelli2020video, baroffio2016camera}, with Mandelli et al.~\cite{mandelli2020video} introducing alignment techniques for stabilized footage. While PRNU has been explored for deepfake detection, it has not yet been applied to deepfake-based fraud in financial contexts. This paper addresses that gap by proposing PRNU as a second-factor authentication method.


\begin{figure}[htbp]
 \centering
\includegraphics[width=0.5\textwidth]{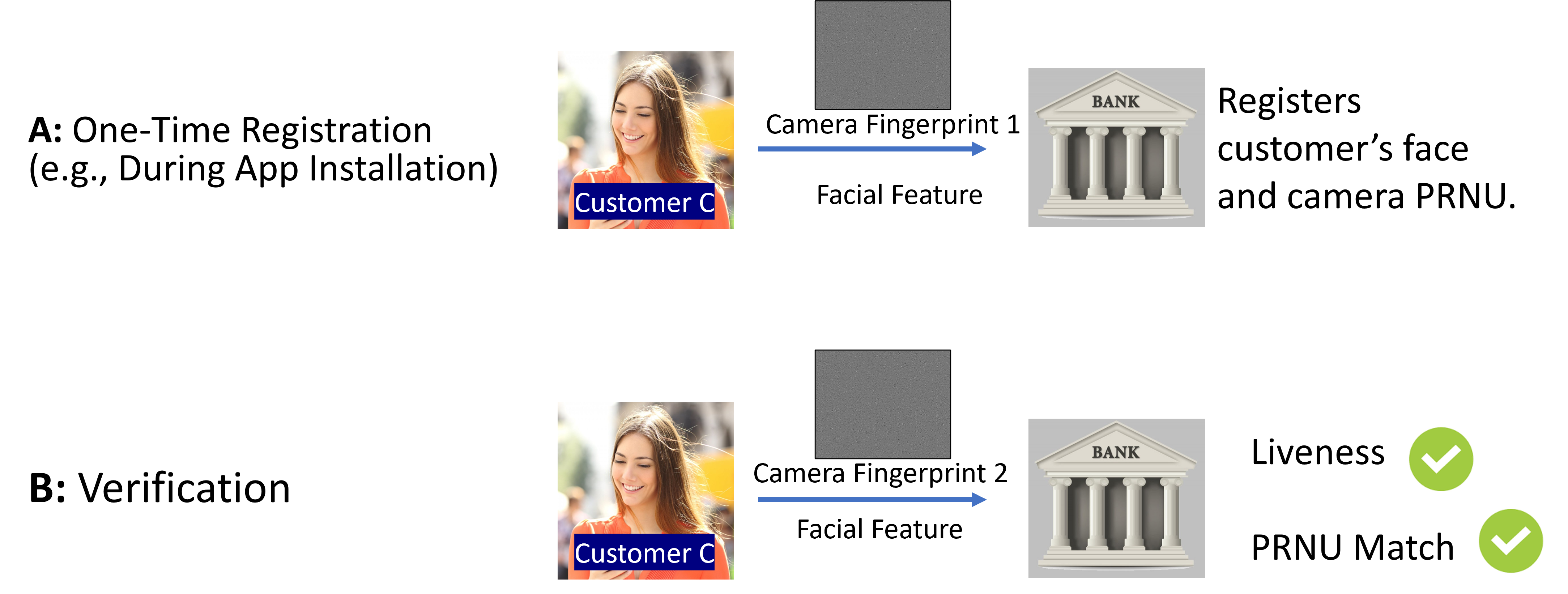}
\caption{Overall idea}
\label{fig3}
\end{figure}

Considering that face-swapping deepfakes change only particular parts of the image this can be taken into account as fraud detection.


\section{Threat Model}
Figure~\ref{fig1} shows the threat model. The bank is a trusted entity that stores customer data securely, e.g., using encryption. An unauthenticated user, such as an attacker, is assumed to be unable to access data from the bank database. The customer is a person who uses bank's service. She is assumed to use selfie banking, and hence authenticates her using her face captured through a selfie camera (e.g., a smartphone selfie camera).  The camera is assumed to be under the control of the customer when an authentication attempt is made by her or on her behalf (i.e., by the attacker). The customer and the bank communicate through a secure network, e.g., using SSL. The attacker is a malicious entity who aims to access a customer's account by leveraging the vulnerability in face-based authentication. It is assumed that the attacker has the required resources to create a deepfake of the customer. The attacker can have the social media images or videos of the customer (that is used to create a deepfake), but the attacker cannot have access to the customer's camera. The attacker can try to log into the customer's account by using the customer's deepfake.

\section{Proposed Method}
We propose using PRNU-based camera attribution as a second-factor authentication method in selfie banking to combat deepfake-based spoofing (Figure~\ref{fig2}). This approach authenticates not only the user’s face but also the camera used, leveraging PRNU without requiring user interaction—unlike OTP-based methods.

As shown in Figure~\ref{fig3}, the method has two stages: (A) Registration and (B) Verification. During registration, the customer’s facial features and camera fingerprint (Fingerprint 1) are captured and securely sent to the bank. The camera fingerprint is computed on the user's device by capturing multiple facial images with the same camera, estimating PRNU noise from each using a denoising filter, and combining them to form a robust fingerprint. This data is stored by the bank for future verification.

\begin{figure}[htbp]
 \centering
\includegraphics[width=0.5\textwidth]{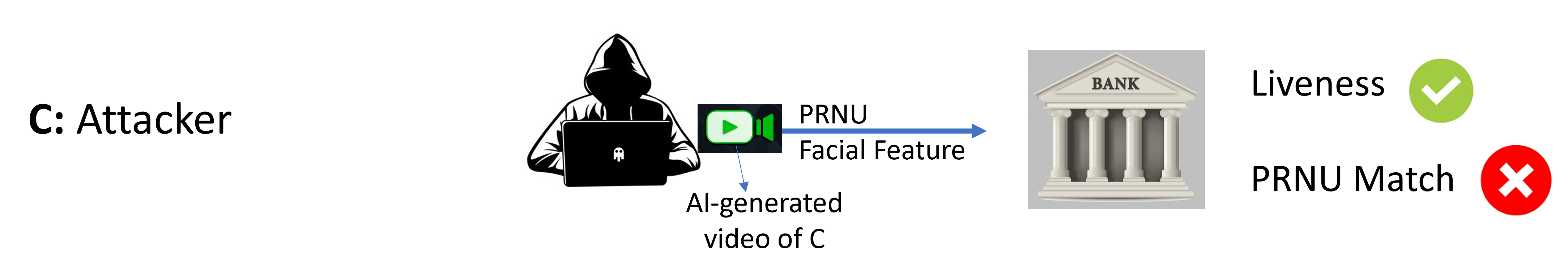}
\caption{Attacker access}
\label{fig4}
\end{figure}
Someone wanting to access the bank is authenticated in the Verification stage using two things: the facial feature and the camera fingerprint. In this stage, fresh facial features and fresh camera fingerprint (i.e., Fingerprint 2) are obtained. These fresh information are sent to the bank in encrypted channels. The bank first does the facial recognition. If this is successful,  the bank does PRNU-based camera authentication. The camera authentication is done by matching the freshly received camera fingerprint (i.e., Fingerprint 2) with the stored fingerprint (i.e., Fingerprint 1). The matching of the fingerprints is done using the PCE method. If the PCE result is above $50$, it is considered that both the fingerprints come from the same camera, and hence the camera authentication is successful. A customer is authenticated only when both face-based authentication and camera authentication are successful. This is because a match implies that (i) the facial features are coming from a camera (not AI generated), and (ii) the camera is registered against the customer. As shown in Figure~\ref{fig4}, the camera-based authentication will fail for a  man-in-middle attacker as she will not have access to the customer's camera. As discussed earlier, it is almost impossible for the attacker to implant a camera fingerprint on a fake image.

\begin{table}[h]
\centering
\begin{tabular}{|c|c|}
\hline
Camera & Images for Fingerprint \\ 
\hline
Dell Inspiron Laptop Camera & 15 \\ 
\hline
Logitech WebCam & 20\\ 
\hline
Samsung A8 Smartphone & 16\\ 
\hline

\end{tabular}
\caption{Cameras used in experiment}
\label{tab:dataset}
\end{table}

\section{Experiments and Results}\label{AA}
Our experiment is done in a simulation of a real-world setup. We considered three different cameras: a smartphone camera, a laptop camera, and a webcam, as a customer can use any one of them in the face-based authentication. Table~\ref{tab:dataset} shows these cameras. The attacker will not have access to these camera, but can access any image from a camera (e.g., a social media image). We therefore tested our scheme using deepfake videos generated using an image from each of the cameras. We then showed that these deepfake videos can defeat the liveness detections but cannot defeat the PRNU tests.

\subsection{Creating deepfakes}
\begin{figure}[htbp]
  \centering
  \begin{minipage}[t]{0.115\textwidth}
    \centering
    \includegraphics[width=\textwidth]{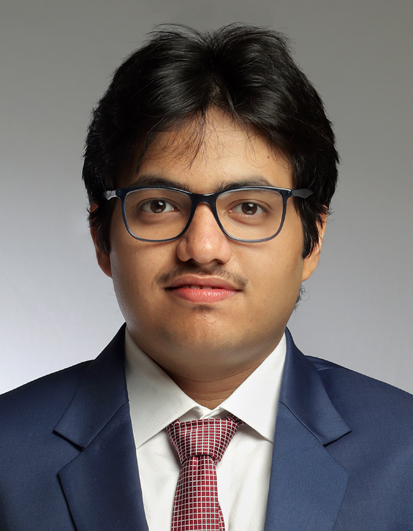}
    \caption*{(a) Source image}
  \end{minipage}
  \hfill
  \begin{minipage}[t]{0.15\textwidth}
    \centering
    \includegraphics[width=\textwidth]{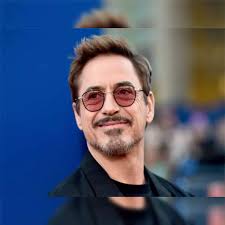}
    \caption*{(b) Destination video frame}
  \end{minipage}
  \hfill
  \begin{minipage}[t]{0.15\textwidth}
    \centering
    \includegraphics[width=\textwidth]{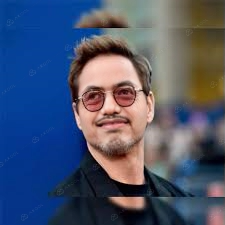}
    \caption*{(c) Deepfake result}
  \end{minipage}
  \caption{Face swapping pipeline: (a) Source image, (b) Frame from destination video, (c) Resulting deepfake.}
  \label{fig:faceswap_row}
\end{figure}

We used the widely available freeware deepfake tool \textit{Akool} for creating a deepfake. For this, we used two inputs: a source image and a destination video. The source image was a photo of one of the authors taken using one of the cameras listed in Table~\ref{tab:dataset} (i.e., one of our authors acted as the customer). The destination video was a publicly available video, such as a celebrity video. Figure~\ref{fig:faceswap_row} shows an example of a source image, a destination image, and a deepfake video. The deepfake video is generated by AKOOL by imposing the source image on the destination video.  

AKOOL works by first computing gradients from source image using convolutional layers for deep learning models followed by regression-based facial alignment networks (FAN). They are then followed by facial segmentation networks using CNNs or GANs. These models output a probability map P (x,y) indicating whether a person belongs to the face. Variational Autoencoders are used to perform the face-swapping experiment, followed by Poisson blending \cite{15056c32-6421-3d30-b07c-412d1d03a0f6} or alpha blending. This is followed by loss functions used in training adversarial loss in case of GANs, and perceptual loss for content similarity.

We performed liveness detection on the deepfake video using an in-house Python code. The code detects liveness by scanning for eye movements, natural eyeball movements, and blinking of the eyes. This scanning was done frame-by-frame, and after $10$ frames, a  result of liveness is obtained. The process is repeated for all frames. At the end, a majority test on all temporary results is done to determine if the liveliness detection is a success or not. Our experiment showed that the deepfake video passed the liveness detection. 

 \subsection{PRNU Test}
The goal here is to show that the deepfake video fails the PRNU test. For this, we first compute a camera fingerprint (i.e., Fingerprint 1) using clear sky images taken by the camera used to obtain the source image (Table~\ref{tab:dataset}). Then a fingerprint from the deepfake video (i.e., Fingerprint 2) was computed. Resizing of the video frames were necessary for bringing the videos to the same resolution before computing fingerprint. Both fingerprints were then matched using PCE. 



\begin{table}[h]
\centering
\begin{tabular}{|c|c|}
\hline
Fingerprint 2 & PCE value \\ 
\hline
Natural image from the laptop camera & 101.559\\ 
\hline
Deepfake from the laptop camera& 45.3992\\ 
\hline
Deepfake from another camera & 3.0243e-04\\ 
\hline

\end{tabular}
\caption{for Laptop}
\label{tab:laptop}
\end{table}

Table~\ref{tab:laptop} shows the result when the laptop camera was used as the camera to take the source image. As shown in the table, the PRNU matching for the deepfake video failed (as PCE < 50), implying that the video did not come from the camera of the customer (i.e., the camera used for Fingerprint 1). The PRNU matching for a deepfake created using a different camera also failed. The PRNU matching, however, was a sucess for a fingerprint computed from a natural image taken by the laptop camera (which means PRNU matching works fine for natural images). 

\begin{table}[h]
\centering
\begin{tabular}{|c|c|}
\hline
Fingerprint 2 & PCE value \\ 
\hline
Natural image from the webcam & 189.1801\\ 
\hline
Deepfake from the webcam & 0.3617\\ 
\hline
Deepfake from another camera & 2.88e-05\\ 
\hline

\end{tabular}
\caption{for WebCam}
\label{tab:Web}
\end{table}

Table~\ref{tab:Web} shows the result when the Webcam was used as the camera to take the source image where as Table~\ref{tab:smart} shows the result when the smartphone was used as the camera to take the source image. The results are similar to the result of the laptop camera. 

\begin{table}[h]
\centering
\begin{tabular}{|c|c|}
\hline
Fingerprint 2 & PCE value \\ 
\hline
Natural image from the smartphone & 2.455e+04\\ 
\hline
Deepfake from the smartphone & -1.635e+04\\ 
\hline
Deepfake from another source & -4.0855e+03\\ 
\hline

\end{tabular}
\caption{for Smartphone}
\label{tab:smart}
\end{table}

In all three tests, we see that deepfake images can be easily singled out and we can easily prevent fraudsters (identity thieves) who use images / videos through large scale deployment of a PRNU based system.

\section*{Conclusion}
In conclusion, while facial recognition has enhanced the usability of authentication in banking through selfie banking, it is increasingly vulnerable to deepfake attacks that can convincingly spoof live faces. Traditional liveness detection methods, though effective against static photos and videos, are not sufficient against AI-generated manipulations that mimic real facial features. To address this, we proposed augmenting facial recognition with PRNU-based source camera authentication as a second factor. By verifying that the image used for authentication originates from a previously registered camera, this method offers a strong defense against deepfake-based spoofing. Experimental results confirm that while deepfakes can bypass liveness detection, they fail the PRNU check, demonstrating the robustness and practical utility of our approach in enhancing the security of biometric authentication systems.

\bibliographystyle{ieeetr}
\bibliography{mybibliography.bib}
\vspace{12pt}

\end{document}